\documentclass[aps, twocolumn]{revtex4-1}

\usepackage{graphicx}
\usepackage{amsmath}
\usepackage{dsfont}
\usepackage{amssymb}
\usepackage{physics}
\usepackage{hyperref}
\usepackage{ulem}
\hypersetup{colorlinks=true,
	    final=true,
	    linkcolor=blue,
	    citecolor=blue,
	    filecolor=blue,
	    urlcolor=blue,}

\usepackage{amssymb}

\newcommand{\rta}{\rightarrow}

\newcommand{\ep}{\epsilon}

\newcommand{\p}{\prime}
\newcommand{\pp}{{\prime\prime}}
\newcommand{\om}{\omega}
\newcommand{\ra}{\rangle}
\newcommand{\la}{\langle}

\newcommand{\beq}{\begin{equation}}
\newcommand{\eeq}{\end{equation}}

\begin{document}

\title{Polarizability of a Wigner crystal}

\author{Navinder Singh}

\email{navinder.phy@gmail.com; navinder@prl.res.in}
\affiliation{Theoretical Physics Division, Physical Research Laboratory (PRL), Ahmedabad, India. PIN: 380009.}

\begin{abstract}
We present a calculation of the imaginary part of the polarizability of a Wigner crystal using the Fluctuation-Dissipation theorem. The oscillations of the localized electrons about their equilibrium positions are treated in the harmonic approximation and the electric dipole-moment -- dipole-moment correlator is computed by a normal mode expansion. The amplitudes and phases of the different normal modes are assumed to be statistically independent. In the first case, polarizability is computed in the high temperature limit, $k_B T>>\hbar \Omega_W$ (here, $\Omega_W$ is the Wigner frequency, analogous to the Debye frequency of the phonon case). In the second case, a general expression (valid both at high and low temperature limits) is obtained using a phenomenological damping model. The connection between our general expression and that of the Lorentz oscillator model is discussed. It turns out that the Wigner crystal would be transparent  for applied frequencies greater than the Wigner frequency. A standard ellipsometry set-up can test the predictions of the theory. 
\end{abstract}

\maketitle

 \section{What is a Wigner crystal?}

 The idea of the "Wigner crystal" originated when Eugene Wigner\cite{wigner1}  pointed out a problem with the argument of Felix Bloch regarding the ferromagnetic instability in metallic systems\cite{bloch1,kubo1}. Felix Bloch had argued\cite{bloch1,kubo1} that ferromagnetism in the iron group metals might be originating from the competition between the ferromagnetic exchange interaction and the electronic kinetic energy. Without exchange interactions, free electrons form a Fermi sphere in momentum space with each momentum state being doubly occupied by two electrons of opposite spin. This is the minimum kinetic energy configuration. 
 
 But in the presence of the ferromagnetic exchange interaction, electrons tend to align their spins. Pauli principle prohibits the presence of two electrons of the same spin in any momentum state. Thus electrons migrate to higher momentum states (which are vacant), and thereby increasing (inflating) the radius of the Fermi sphere (or enhancing the total kinetic energy). But it {\it lowers} the ferromagnetic exchange energy. If ferromagnetic exchange energy wins the competition then the system transitions to a ferromagnetic instability. [note that this is before Stoner\cite{stoner}].
 
It turns out that parallel spin electrons avoid each other in the real space too, because by doing so there is a net reduction in the electrostatic Coulomb repulsion (the actual mechanism behind the ferromagnetic exchange interaction)\cite{bloch1,kubo1,navhist}. Bloch argued that if the electron density is sufficiently low, this ferromagnetic instability might occur, and these are parallel spin electrons which leads to exchange interaction lowering of energy. He came up with the inequality:
\beq
4.5 \frac{e^2 m^*}{h^2}> n^{1/3}.
\eeq
Here, $m^*$ is electronic effective mass and $n$ is the electron density. {\it However, Eugene Wigner pointed out\cite{wigner1} that the Coulomb correlation energy reduction must happen not only between two electrons with parallel spins, but also between two electrons with opposite spins (Bloch had neglected the Coulomb correlation between electrons of opposite spins)}.  Wigner argued that if correlation energy is taken into account (for both, the opposite spins pairs and the same spin pairs) then Bloch's criterion is too restrictive, and instead, the system of electrons (with sufficiently low density) might actually crystallize in real space, rather than transitioning into a ferromagnetic instability. 
 
 \begin{figure}[h!]
    \centering
    \includegraphics[width=1.0\columnwidth]{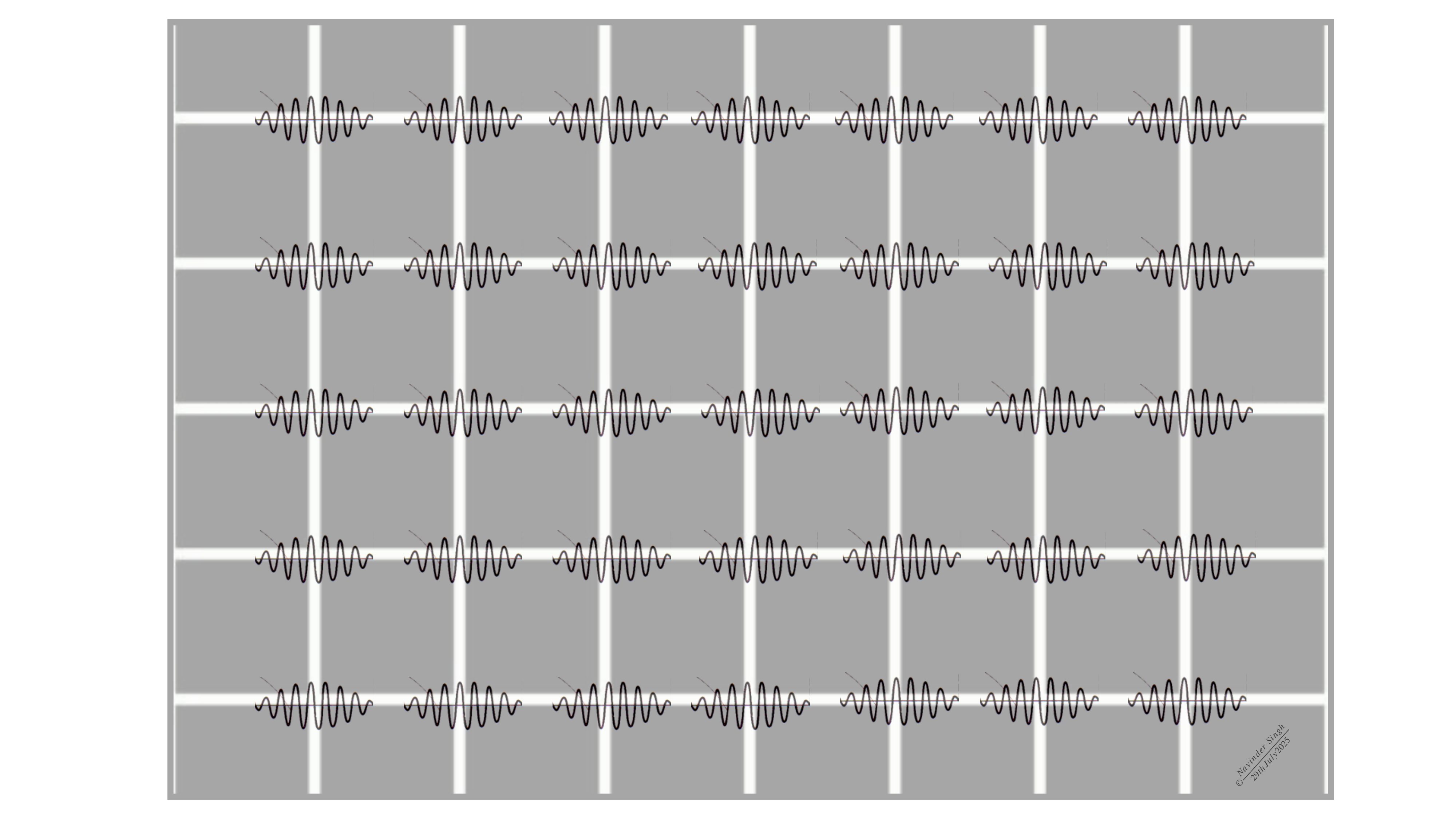}
    \caption{A cartoon of a 3D Wigner crustal (top view). Electrons are depicted as wave-packets localized around the lattice sites (this is called the "on-site" Wigner crystal).}
  \end{figure}
 
To clarify the argument, Wigner considered an extreme limit: no kinetic energy situation\cite{wigner1}. Then, whether the spins are parallel or antiparallel, electrons would settle in a minimum potential energy configuration which is achieved if they form a crystalline lattice in real space (avoiding each other in real space (figure 1)). In Wigner's words\cite{wigner1}:
 
 \begin{quote}
{\texttt{``If the electrons had no kinetic energy, they would settle in configurations which correspond
to the absolute minima of the potential energy. These are closed-packed lattice configurations,
with energies very near to that of the body centered lattice. Here, every electron is very nearly surrounded with a spherical hole of
radius $r_s$ and the potential energy is smaller than in the random configuration by the amount $0.75= \frac{e^2}{r_s}$. 
This would be the sum of the correlation energy and that due to the Fermi hole.''  -- Eugene Wigner.}}
\end{quote} 

The crucial role is played by the {\it Fermi hole} (an interesting and insightful discussion of the Fermi hole and its role in context of exchange contributions in the Hartree-Fock approximation is given in section 6.2 in\cite{ds}). In the case of the finite kinetic energy, this instability (real apace crystallization of electrons) can occur if kinetic energy (K.E.) is much smaller than the potential energy (P.E.). In other words, the parameter\cite{setsuo}:

\beq
\Gamma = \frac{P.E.}{K.E.},
\eeq

must be much greater than one for a Wigner crustal (WC) to form. A rough expression for $\Gamma$ can be made in this manner\cite{setsuo}. If electrons are separated from each other by an average distance $a$, then P.E. per electron would be $\frac{e^2}{4\pi\ep_0 a}$. And the K.E. energy would be $\frac{\hbar^2 k_F^2}{2 m} \sim \frac{\hbar^2}{2 m a^2},~~~k_F \sim\frac{1}{a}$. Then the above mentioned ratio (let us say in 2D with $a\sim n^{-1/2}$) scales as

\beq
\Gamma \sim \frac{m}{\sqrt{n}}.
\eeq
Where $n$ is the electron density. At sufficiently low density (thus high $\Gamma$) this transition can occur. Accurate numerical calculations\cite{tanatar} give a value $\Gamma_c \sim 34$. If $\Gamma$ is greater than $\Gamma_c$ this transition can occur.

There is already a body of evidence (although indirect) of WC formation in $GaAs/Al_xGa_{1-x}As$ hetro-structures\cite{ev1,ev2,ev3}. Electron density in such hetro-structures can be controlled by gate voltage and effective electronic mass can be controlled by the application of magnetic fields (through flatter band formation).

Very recently (in 2024), there is a {\it direct} evidence of WC formation in Bernal-staked bilayer graphene\cite{tsui}. This is enabled by very high resolution STM imaging techniques\cite{tsui}. In Moire hetro-structures also, under suitable conditions, there is an evidence of WC formation. Interested readers can refer to a review article\cite{rev1}, for further details and references to the modern literature therein.

In this article, we present a calculation of the imaginary part of the dielectric polarizability ($\alpha^\pp(\om)$) of a WC using the Fluctuation-dissipation theorem.  This will be useful for an experimental characterization of the Wigner crystal (i.e., experimentally determining its Wigner frequency etc) and leads to a deeper insight into its dissipative mechanisms and excitations of Wigner crystals.  The oscillations of the localized electrons about their equilibrium positions are assumed small, and thus can be treated in the harmonic approximation.  The calculation is divided into two cases:  (1) $\alpha^\pp(\om)$ is computed in the high temperature limit $k_B T>>\hbar \Omega_W$ (where $\Omega_W$ is the 'Wigner frequency' very much analogous to the Debye frequency of the lattice case). We find that in this regime $\alpha^\pp(\om)$ scales linearly with $\om$. (2) In the second case, we obtain a general expression for $\alpha^\pp(\om)$ (which is valid both at  the high temperature limit and the low temperature limit) using a phenomenological damping model. The connection between our general expression and that of the standard Lorentz oscillator model of insulators is discussed.

This paper is organized in the following manner: Section (II) has two sub-sections. In II(A) $\alpha^\pp(\om)$ is computed in the high temperature limit, and in II(B) we introduce a phenomenological damping model. Physical reasoning of damping and results are discussed in the end of section (II). A summary and conclusion is presented in section (III).

\section{Calculation of the dielectric polarizability of a Wigner crystal}

\subsection{Dissipation due to stochasticity in the high temperature limit ($k_B T>>\hbar \om_W$)}

The imaginary part of the dielectric polarizability of the Wigner crystal  can be computed using the Fluctuation-Dissipation theorem (FDT)\cite{kubo2,navb}:
\beq
\alpha^\pp(\om)  = \frac{1}{2\hbar} (1-e^{-\beta\hbar\om})\int_{-\infty}^{+\infty} dt e^{i\om t} \la (-e) x(t) (-e) x(0)\ra,
\eeq 
where $-e x(t)$ is the electric dipole moment operator. The above equation can be written as:
\beq
\alpha^\pp(\om)  = \frac{e^2}{2\hbar} (1-e^{-\beta\hbar\om})\int_{-\infty}^{+\infty} dt e^{i\om t} \la x(t) x(0)\ra
\eeq 

The central object is the position-position correlator ($\la x(t) x(0)\ra$). It is computed in ref.\cite{nav10} for lattice vibrations (phonon case).  Basically,  here we consider a 3D Wigner crystal with electrons localized in simple cubic symmetry with lattice constant $a$. The electrons are localized just above the ions (this is called the 'on-site' Wigner crystal). Any slight deviations $x(t)$ from their equilibrium positions would lead to an induced electric dipole moment $-e x(t)$. Consider longitudinal acoustic  vibrations (very much analogous to acoustic lattice vibrations (phonon case)) such that the wave vector $q$ is along the lattice constant vector $a$ joining the two nearby electrons.  It is appropriate to call them as electronic acoustic vibrations (acoustic modes). The instantaneous distance between any two nearby electrons $x(t)$ can be written in terms of the normal modes of vibration\cite{nav10}: 
\beq
x(t) = \frac{1}{N^{3/2}}\sum_q A_q\cos(\Omega_q t +\phi_q)\cos(qa).
\eeq
Here $\om_q,~A_q,~\phi_q$ are the frequency, amplitude, and phase of the normal mode defined by wave vector $q$, respectively. $N$ is the number of unit cells with volume of each unit cell $a^3$. $A_q$ and $\phi_q$  are taken to be uncorrelated quantities for different modes. From equation (6), one can write

\begin{eqnarray}
\la x(t_1)x(t_2)\ra &=& \frac{1}{N^3}\sum_{q,q'}\la A_q A_{q'}\ra \cos(qa)\cos(q'a)\nonumber\\
&\times&\la\cos(\Omega_q t_1+\phi_q)\cos(\Omega_{q'}t_2 + \phi_{q'})\ra,
\end{eqnarray}
where different modes are treated independent from each other, such that $\la A_q A_{q^\p}\ra = \la A_q^2\ra \delta_{q,q^\p}$. We obtained\cite{nav10}:

\beq
\la x(t)x(0)\ra = \frac{a^3\hbar}{6\pi^2m c_s^3} \int_0^{\Omega_W} d\Omega \frac{\Omega}{e^{\beta \hbar \Omega} -1} \cos(\Omega t)\cos^2(\Omega t_s).
\eeq

Here $\Omega_W$ is an analogue of the Debye frequency (call it the Wigner frequency). Define a lattice time scale: $t_s = a/c_s$. Where $c_s$ is the sound speed (propagated by the normal modes of the localized electrons) in the Wigner crystal.  Consider the frequency integral (in equation (8)): 

\beq
I = \int_0^{\Omega_W} d\Omega \underbrace{ \frac{\Omega}{e^{\beta \hbar \Omega}-1} \cos^2(\Omega t_s)}_{Ist}\underbrace{\cos(\Omega t)}_{2nd}.
\eeq

In the high temperature limit $k_BT>>\hbar\Omega_W$, the above expression simplifies to:

\beq
I \simeq \frac{k_B T}{\hbar} \cos^2(\Omega_W t_s) \left(\frac{\sin\Omega_W t}{t}\right).
\eeq
In the considered limit, the other part is a sub-leading term whose magnitude is much smaller than the leading term, and thus can be neglected. Further, if the time-scale of the measurements is much greater than the inverse of the Wigner frequency $t>>\frac{1}{\Omega_W}$, then we can use:

\beq
\frac{\sin(\Omega_W t)}{t} \rta \pi \delta(t).
\eeq
Therefore, the expression for $\alpha^\pp(\om)$ (equation (5)) is: 

\begin{widetext}
\beq
\alpha^\pp(\om) = \frac{e^2}{12 \pi m \om_W^2} (a q_W)^3 \cos^2(a q_W) \left[\frac{k_BT}{\hbar\Omega_W}\right] (1-e^{-\beta\hbar\om}) \left( \int_{-\infty}^{+\infty}dt e^{i \om t}\delta(t)\right).
\eeq
\end{widetext}

The important point to be noted is that the timescale involved in the integral must be much much greater than the inverse of the Wigner frequency. That is, if $1/\om_W$ is of the order of $femtoseconds$, then the time scale related to external measurements must be in the pico-second scale, such that $\Omega_W t>>1$ and equation (11) holds true. This implies that the applied frequency for measurement of $\alpha^\pp(\om)$ must be much less that the characteristic Wigner frequency ($\om<<\Omega_W$) as the associated timescale with it is in pico-seconds. Physically speaking, in this low frequency regime, the average behaviour of the correlator will exhibit stochasticity in the form of delta correlation in time (this argument is thoroughly discussed in\cite{nav10}).  

One can estimate the Wigner frequency from the expression $\Omega_W = c_s q_W$. If we take $q_W \sim \frac{1}{a}, ~a = 10\AA$, and $c_s = 10^{5} m/sec$ (considering that electron mass is three orders of magnitude smaller than the ionic mass, and typical sound speed (for phonons) is $10^{3} m/sec$), we get $\Omega_W \sim 10^{14} ~Hz$. Thus, experiments done at micro-wave and infra-red frequencies satisfies the criterion. The expression of the imaginary part of the dielectric polarizability (equation (12)) is given by

\beq
\alpha^\pp(\om) = \frac{e^2}{12 \pi m \Omega_W^2} (a q_W)^3 \cos^2(a q_W) \left[\frac{\om}{\Omega_W}\right],
\eeq

or

\beq
\alpha^\pp(\om) \sim \frac{\om}{\Omega_W}.
\eeq

However, this is valid only in the limit $k_BT>>\hbar\Omega_W>>\hbar\om$. This is a very restrictive condition. The reason is the emergence of the dissipative character in the system only in the high temperature limit\cite{footnote1}. Below we introduce a phenomenological model (which is applicable in general) in which the dissipative character is introduce at the very beginning, but it is introduced ad-hoc.

\subsection{A phenomenological dissipation model}

To rectify the issue of the very limited regime of applicability of the above derived expression for $\alpha^\pp(\om)$, we  introduce a phenomenological model. We assume that each normal mode decay with time with a decay constant which is in fact wave-vector dependent. Physical reasoning why it happens is given at the end of this section. We assume that the amplitudes follow: 

\beq
A_q \rta A_q e^{-\Gamma_q t/2}.
\eeq

Here $\Gamma_q$ is a mode dependent damping constant. With this assumption, equation (7) with $t_1=t,~t_2=0$ leads to 
\beq
\la x(t)x(0)\ra = \frac{1}{N^3}\sum_{q}\la A^2_q\ra \cos^2(qa)\cos(\om_q t) e^{-\Gamma_q t}.
\eeq
Again, the different modes are assumed to be uncorrelated. Changing summation into integration we again get equation (8) but now with an extra factor of $e^{-\Gamma_q t}$. The same expression (the above equation) is also obtained for $\la x x(t)\ra$ when we set $t_1 =0$ and $t_2=t$ in equation (7). The full Fourier transform in equation (5) can be split into two half-Fourier transforms as

\beq
\int_{-\infty}^{+\infty} dt e^{i\om t} \la x(t)x\ra = \int_0^\infty dt \la x(t)x\ra + \int_0^\infty dt e^{-i\om t}\la x x(t)\ra,
\eeq
where we changed the limits in the negative part as $t\rta -t$ and used the cyclic properties of the trace: $\la x(-t) x\ra= tr(e^{\beta H}x(-t)x) = tr(e^{\beta H}e^{-i H t} x e^{i Ht} x) = tr(e^{\beta H} x e^{i Ht} x e^{-i H t}) = \la x x(t)\ra$. Here $H$ is the Hamiltonian of the Wigner crystal $H = \sum_q \hbar \om_q (n +1/2)$. And the operators are in the Heisenberg representation. Using equations (8),  (16) and (17), the imaginary part of the polarizability (equation (5)), after lengthy but straightforward calculation, turns out to be: 
\begin{widetext}
\beq
\alpha^\pp(\om) = \frac{e^2}{12 \pi m \Omega_W^3} (a q_W)^3 \cos^2(a q_W) (1-e^{-\beta\hbar\om})\int_0^{\Omega_W} d\Omega  (\Omega^2 +\om^2) \left( \frac{1}{e^{\beta\hbar\Omega} -1} +\frac{1}{2} \right)\left( \frac{1}{\pi}  \frac{2 \Omega \Gamma(\Omega)}{(\om^2-\Omega^2)^2 + 4\om^2 \Gamma(\Omega)^2}  \right)
\eeq
\end{widetext}

This is our main result. We set $\Gamma_q = \Gamma(\Omega)$ with $\Omega= c_s q$, and $\Omega_W = c_s q_W$. And $q_W$ is the Wigner wave vector. It is completely general in the sense that once the line-width variation with frequency $\Gamma(\Omega)$ of the acoustic modes is known, the imaginary part of the polarizability can be computed for a given value of applied frequency, temperature, and the characteristic Wigner frequency of the Wigner crystal. In this way, the expression (18) can be combined with  ab-initio calculations like an appropriate modification/generalization of DFT (density functional theory) appropriate for  the localized electrons in the Wigner crystal. From such a "Wigner DFT" setting, acoustic modes of electrons (electronic phonon modes) and their broadening ($\Gamma(\Omega)$, by including anharmonic processes) can in principle be calculated, and detailed studies (material specific) can be done\cite{dft1,dft2}.  It is to be noted that the above expression goes to equation (13) in the limit $\Gamma(\Omega)\rta0$, and $k_BT>>\hbar\Omega_F>>\hbar\om$. This is a consistency check. 

However, let us consider a simplified case. If we assume that $\Gamma(\Omega) = \Gamma$ (that is, the constant damping case) and consider the zero temperature limit ($T\rta 0$), and further set $q_W = \frac{2\pi}{a}$, we get a simplified zero temperature limit expression: 

\beq
\alpha^\pp(\om) = \frac{2\pi e^2 \Gamma}{3 m \Omega_W^3}\int_0^{\Omega_W} d\Omega \frac{\Omega (\om^2 + \Omega^2)}{(\om^2-\Omega^2)^2 +4 \om^2\Gamma^2}.
\eeq

In the very high frequency limit $\om>>\Omega_W$, the imaginary part of the polarizability (from the above expression) scales as $1/\om^2$. This is to be contrasted with the high frequency limit behaviour of $\alpha^\pp(\om)$ from the Lorentz oscillator model of insulators (equation (21)) where it scales as $1/\om^3$.

It is useful to compare the main result (equation (18)) with the Lorentz oscillator model. Dielectric polarizability can be computed using the Lorentz model for insulators\cite{navb}. In the Lorentz model, each localized electron is assumed to be independently oscillating about its equilibrium position with a characteristic frequency (interaction between any two oscillating electrons is neglected). If each electron is assumed to be confined in a harmonic potential then its Hamiltonian is given by $\hbar\Omega_0(n+1/2)$ where $\Omega_0$ is the fundamental frequency of the oscillator. The equation of motion of the oscillating electron is that of the damped harmonic oscillator :

\beq
m\frac{d^2 x(t)}{dt^2} + m x(t)\Omega_0^2 = -eE -m\Gamma \frac{dx(t)}{dt}.
\eeq

Electric dipole moment at time $t$ is given by $-e x(t)$. The total induced dipole moment is proportional to the applied electric field: $P = \alpha(\om)E$, and the imaginary part of the polarizability turns out to be\cite{navb}: 

\beq
\alpha^\pp(\om) = \frac{e^2}{m}\frac{\om\Gamma}{(\om^2-\Omega_0^2)^2 +(\om\Gamma)^2}.
\eeq

The connection between the Lorentz oscillator model  (equation (21)) and our phenomenological model (18) is also clear.  Basically, equation (18) is a thermally weighted sum of a continuous spectrum of Lorentz oscillators located at frequencies $\Omega$ instead of $\Omega_0$, the whole is integrated from 0 to $\Omega_W$.  In the $\om\rta 0$ limit, both the expressions (equations (18) and (21)) leads to $\alpha^\pp(\om) \sim \om$ (that is, linear increase in frequency in the low frequency limit). 

A numerical plot of equation (18) by considering constant damping case is given in figure (2).
 \begin{figure}[h!]
    \centering
    \includegraphics[width=1.0\columnwidth]{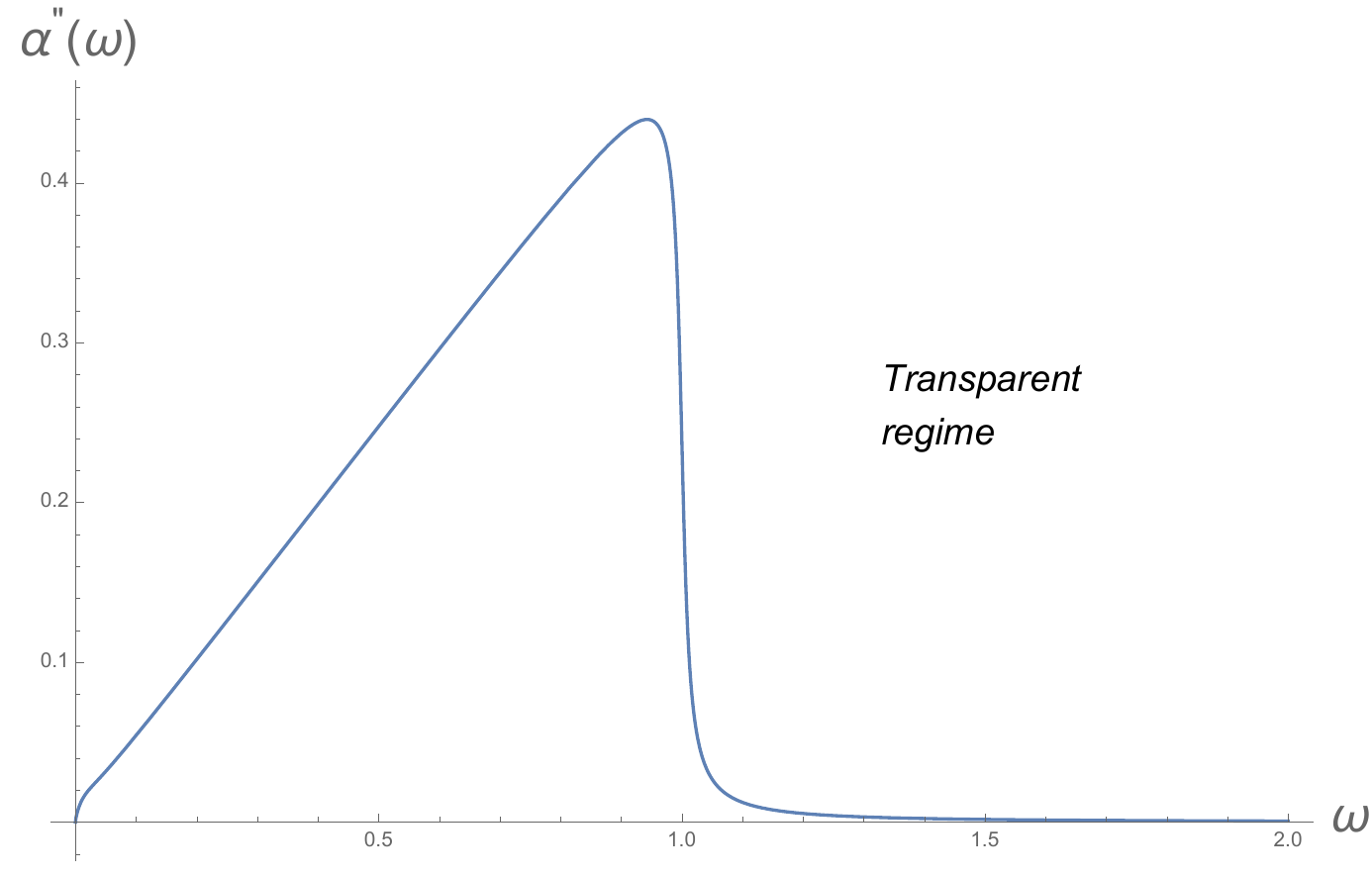}
    \caption{The imaginary part of the polarizability as function of frequency. For the numerical calculation we set: $\Omega_W = 1~eV,~~T= 100~K $, and constant damping $\Gamma = 0.01~eV$.  For $\om>\Omega_W$ the imaginary part of the polarizability suddenly drops to very small values implying negligible dissipation inside the Wigner crystal. The imaginary part of the dielectric function (which is proportional to the imaginary part of the polarizability) also becomes very small implying sudden transparency of the Wigner crystal for this high frequency limit.}
  \end{figure}
For $\om>\Omega_W$ the imaginary part of the polarizability decays to very small values implying transparency of the Wigner crystal in this high frequency limit.  In the low frequency limit $\om<\Omega_W$, $\alpha^\pp(\om)$ roughly linearly increases with frequency  (the values of temperature, damping ($\Gamma$), and $\Omega_W$ used to plot figure (2) are given in the figure caption).

We come back to the question of the physical reasoning of the damping (equation (15)). As in the standard phonon-phonon (anharmonic) interactions in metals, a given phonon can decay by merging (coalescing) with already existing phonon, or it can decay by splitting itself into two phonons of lower energy. That is, a phonon of energy $\hbar \om_0$ can split into two phonon of energies $\hbar\om_1$ and $\hbar\om_2$, such that $\hbar \om_0 = \hbar\om_1+\hbar\om_2$. A merger (coalescing) of two phonons can also happen $\hbar\om_1+\hbar\om_2 \rta \hbar\om_0$. Looking at equation (15) this interpretation seems to not making any sense.  However, if we examine equation (16) we notice that the damping factor $e^{-\Gamma_q t}$ is multiplied with $\la A_q^2\ra$. Now, for the harmonic oscillator case (which we are considering) we have 
\beq
\la A_q^2\ra = \frac{\hbar}{3 m \Omega_q} \left(n_q +\frac{1}{2}\right).
\eeq
Where $n_q =\frac{1}{e^{\beta\hbar\Omega_q} -1}$. Thus, the updated expression of $\la A_q^2\ra$ will have a term $n_q e^{-\Gamma_q t}$ which means that the number of phonons in mode $q$ decays with time. The rate of this decay is given by $\Gamma_q$. In this way the physical meaning of the damping can be understood. However, it must be remembered that the energy scale associated with the damping rate must be much smaller than the mode frequency (in fact, $ \Gamma_q << \Omega_q$) so that the anharmonic processes acts just like a very small perturbation. This is our basic underlying assumption which is quite reasonable (anharmonic terms are generally smaller in magnitude than the harmonic terms\cite{dft1}).

We comment that the characteristic behaviour depicted in figure (2) can be experimentally verified. The standard ellipsometry set-up can be used to measure frequency dependent reflectivity of a Wigner crystal. Through the standard Kramers-Kronig analysis, the real and imaginary parts of the dielectric function can be computed. From the imaginary part of the dielectric function, the imaginary part of the polarizability can be determined\cite{navb} and can be compared with the plot shown in figure (2).

\section{Conclusion}

Expressions (equations (13), (14), (18), and (19)) for the imaginary part of the polarizability of an on-site Wigner crystal are derived from the fluctuation-dissipation theorem and these are our main results. Expressions (13) and (14) are valid in the high temperature limit ($\hbar\Omega_W>>k_BT$) whereas the expression (18) has a general validity and it can be seen as a generalization of the Lorentz oscillator model for insulators.  The numerical plot (from the expression (18)) shows that $\alpha^\pp(\om)$ increases roughly linearly with frequency up to what is called the Wigner frequency ($\Omega_W$). For frequencies greater than the Wigner frequency, $\alpha^\pp(\om)$ exhibits a sudden drop implying the transparency of the Wigner crystal (much like UV transparency in metals). We suggest that the standard ellipsometry experimental set-up can test this behavior of $\alpha^\pp(\om)$ for a given Wigner crystal.



\end{document}